\documentclass[5p,twocolumn]{elsarticle}

\usepackage{amsmath,amssymb}
\newcommand{\eqdef}{\stackrel{\text{def}}{=}}
\newcommand{\n}{\nonumber \\}
\newcommand{\bm}{\boldsymbol}
\newcommand{\ignore}[1]{}

\allowdisplaybreaks[4]

\journal{Physics Letters B}

\begin{document}

\begin{frontmatter}

\title{Infinitely many shape invariant potentials and
new orthogonal polynomials}

\author[SO]{Satoru Odake\corref{cSO}}
\ead{odake@azusa.shinshu-u.ac.jp}
\author[RS]{Ryu Sasaki}
\address[SO]{Department of Physics, Shinshu University,
     Matsumoto 390-8621, Japan}
\address[RS]{Yukawa Institute for Theoretical Physics,
     Kyoto University, Kyoto 606-8502, Japan}
\cortext[cSO]{Corresponding author.}

\begin{abstract}
Three sets of exactly solvable one-dimensional quantum mechanical
potentials are presented.
These are shape invariant potentials obtained by deforming the radial
oscillator and the trigonometric/hyperbolic P\"oschl-Teller potentials
in terms of their degree $\ell$ polynomial eigenfunctions.
We present the entire eigenfunctions for these Hamiltonians
($\ell=1,2,\ldots$) in terms of new orthogonal polynomials.
Two recently reported shape invariant potentials of Quesne and
G\'omez-Ullate et al.'s are the first members of these infinitely
many potentials.
\end{abstract}




\end{frontmatter}

\section{Introduction}
\label{intro}
\setcounter{equation}{0}

In this Letter we present two infinite sets and one finite set of
exactly solvable one-dimensional quantum mechanical Hamiltonians.
As the main part of the eigenfunctions, a new type of orthogonal
polynomials is obtained for each Hamiltonian.
They are exactly solvable by combining shape invariance \cite{genden}
with the factorisation method \cite{infhul,crum} or the so-called
supersymmetric quantum mechanics \cite{susyqm}.
Then the entire energy spectrum and the corresponding eigenfunctions
can be obtained algebraically.
However, these new shape invariant Hamiltonians do not possess
the exact Heisenberg operator solutions \cite{os7}, in contrast to
most of the known shape invariant Hamiltonians.

Shape invariance is a sufficient condition for exactly solvable
quantum mechanical systems. Based on one shape invariant potential,
an infinite number of exactly solvable potentials and their
eigenfunctions can be constructed by a modification of Crum's method
\cite{adler,dubov}.
But these newly derived systems fail to inherit the shape invariance, 
nor do they possess Heisenberg operator solutions. Although several
shape invariant `discrete' quantum mechanical systems are added to
recently \cite{os4}, the catalogue of the shape invariant potentials
was rather short for a long time.
In 2008, Quesne \cite{quesne} reported two new shape invariant
potentials based on the Sturm-Liouville problems for the $X_1$-Laguerre
and the $X_1$-Jacobi polynomials proposed by G\'omez-Ullate et al.
\cite{gomez}.

Here we present our preliminary results on the three sets of shape
invariant potentials and the corresponding new types of orthogonal
polynomials, without proof. After brief introduction of notation and
the shape invariance method, they are obtained by {\em deforming\/}
the well-known shape invariant potentials, the radial oscillator and
the Darboux-P\"oschl-Teller \cite{darboux, PT} potentials, in terms
of the degree $\ell$ polynomial eigenfunctions, {\em i.e.\/} the
Laguerre and the Jacobi polynomials. The eigenpolynomials of the new
Hamiltonians are orthogonal polynomials starting from degree $\ell$,
which could be called $X_{\ell}$ polynomials. The Quesne-G\'omez-Ullate
et al. examples \cite{quesne,gomez} correspond to the $\ell=1$ cases.

\section{General setting: shape invariance}
\label{settings}

The starting point is a generic one-dimensional quantum mechanical
system having a square-integrable groundstate together with a finite
or infinite number of discrete energy levels:
$0=\mathcal{E}_0 <\mathcal{E}_1 < \mathcal{E}_2 < \cdots$.
The groundstate energy $\mathcal{E}_0$ is chosen to be zero, by
adjusting the constant part of the Hamiltonian.
The {\em positive semi-definite\/} Hamiltonian is expressed in a
factorised form \cite{infhul,crum,susyqm}:
\begin{align}
  \mathcal{H}&=\mathcal{A}^{\dagger}\mathcal{A}=p^2+U(x),\quad
  p=-i\partial_x,\\
  \mathcal{A}&\eqdef\partial_x-w^{\prime}(x),\quad
  \mathcal{A}^{\dagger}=-\partial_x-w^{\prime}(x),\\
  &\qquad
  U(x)\eqdef w^{\prime}(x)^2+w^{\prime\prime}(x).
\end{align}
For simplicity of presentation we have adopted the unit system in
which $\hbar$ and the mass $m$ of the particle are such that $\hbar=2m=1$.
Here we call a real and smooth function $w(x)$ a {\em prepotential\/}
and it parametrises the groundstate wavefunction $\phi_0(x)$,
which has {\em no node\/} and can be chosen real and positive,
$\phi_0(x)=e^{w(x)}$.
It is trivial to verify $\mathcal{A}\phi_0(x)=0$ and
$\mathcal{H}\phi_0(x)=0$.

{\em Shape invariance\/}, a sufficient condition for exact
solvability \cite{genden}, is realised by specific dependence of
the potential, or the prepotential on a set of parameters
$\bm{\lambda}=(\lambda_1,\lambda_2,\ldots)$, to be denoted by
$w(x;\bm{\lambda})$, $\mathcal{A}(\bm{\lambda})$,
$\mathcal{H}(\bm{\lambda})$, $\mathcal{E}_n(\bm{\lambda})$, etc.
The shape invariance condition to be discussed in this Letter is
\begin{align}
  &\mathcal{A}(\bm{\lambda})\mathcal{A}(\bm{\lambda})^{\dagger}
  =\mathcal{A}(\bm{\lambda}+\bm{\delta})^{\dagger}
  \mathcal{A}(\bm{\lambda}+\bm{\delta})
  +\mathcal{E}_1(\bm{\lambda}),
  \label{shapeinv1}\\
  &w^{\prime}(x;\bm{\lambda})^2
  -w^{\prime\prime}(x;\bm{\lambda})\n
  &=w^{\prime}(x;\bm{\lambda}+\bm{\delta})^2
  +w^{\prime\prime}(x;\bm{\lambda}+\bm{\delta})+\mathcal{E}_1(\bm{\lambda}),
  \label{shapeinv2}
\end{align}
in which $\bm{\delta}$ is a certain shift of the parameters.
Then the entire set of discrete eigenvalues and the corresponding
eigenfunctions of $\mathcal{H}=\mathcal{H}(\bm{\lambda})$
\begin{equation}
  \mathcal{H}(\bm{\lambda})\phi_n(x;\bm{\lambda})
  =\mathcal{E}_n(\bm{\lambda})\phi_n(x;\bm{\lambda})
\end{equation}
is determined algebraically \cite{genden,susyqm,os4}:
\begin{align}
  \mathcal{E}_n(\bm{\lambda})
  &=\sum_{k=0}^{n-1}\mathcal{E}_1(\bm{\lambda}+k\bm{\delta}),\\
  \phi_n(x;\bm{\lambda})&\propto
  \mathcal{A}(\bm{\lambda})^{\dagger}
  \mathcal{A}(\bm{\lambda}+\bm{\delta})^{\dagger}\cdots
  \mathcal{A}(\bm{\lambda}+(n-1)\bm{\delta})^{\dagger}\n
  &\qquad\qquad\qquad\qquad\qquad \times e^{w(x;\bm{\lambda}+n\bm{\delta})}.
  \label{genformula}
\end{align}

\section{The radial oscillator}
\label{radialosci}

Here we present an infinite number of shape invariant potentials
indexed by a non-negative integer $\ell=0,1$, $2,\ldots$.
For $\ell=0$, it is the well-known radial oscillator, or the harmonic
oscillator with a centrifugal barrier potential, with $\bm{\lambda}=g>0$:
\begin{align}
  &\mathcal{H}_0(g)=p^2+x^2+\frac{g(g-1)}{x^2}-1-2g,\\[-2pt]
  &w_0(x;g)=-\tfrac12 x^2+g\log x, \quad 0<x<\infty.
\end{align}
Here we adopt the notation of our previous work \cite{os7} \S{III.A.1}.
The shape invariance, the Heisenberg operator solution and the
creation-annihilation operators of the above Hamiltonian are discussed
in some detail there.
It is trivial to verify \eqref{shapeinv2} with $\bm{\delta}=1$,
$\mathcal{E}_1(g)=4$ and we obtain the equidistant spectrum and the
corresponding eigenfunctions $n=0,1,2,\ldots$,
\begin{align}
  &\mathcal{E}_n(g)=4n,\\
  &\phi_n(x;g)=P_n(x^2;g)\,e^{w_0(x;g)},
  \,P_n(x;g)=L_n^{(g-\frac12)}(x).\!\!\!
\end{align}
The polynomial eigenfunctions are the Laguerre polynomials in $x^2$,
which are orthogonal with respect to the measure
$\phi_0(x)^2=e^{2w_0(x;g)}=e^{-x^2}x^{2g}$.

For each positive integer $\ell\ge1$, let us introduce a prepotential
and a Hamiltonian:
\begin{align}
  &\xi_{\ell}(x;g)\eqdef L_{\ell}^{(g+\ell-\frac32)}(-x),
  \label{xilag}\\
  &w_{\ell}(x;g)\eqdef w_0(x;g+\ell)
  +\log\frac{\xi_{\ell}(x^2;g+1)}{\xi_{\ell}(x^2;g)},
  \label{wllag}\\
  &\mathcal{A}_{\ell}(\bm{\lambda})\eqdef\partial_x
  \!-w_{\ell}'(x;\bm{\lambda}),
  \ \mathcal{A}_{\ell}(\bm{\lambda})^{\dagger}
  =-\partial_x\!-w_{\ell}'(x;\bm{\lambda}),
  \label{Aldef}\\
  &\mathcal{H}_{\ell}(\bm{\lambda})\eqdef
  \mathcal{A}_{\ell}(\bm{\lambda})^{\dagger}\mathcal{A}_{\ell}(\bm{\lambda}).
  \label{Hldef}
\end{align}
Since the polynomial $\xi_{\ell}(x^2;g)$ has no zero in the domain
$0<x<\infty$, the prepotential and the potential are smooth in the
entire domain. It is straightforward to verify the shape invariance
condition \eqref{shapeinv2} with
$\bm{\delta}=1$, $\mathcal{E}_{\ell,1}(g)=4$.
By using \eqref{genformula} as a Rodrigues type formula,
we obtain the complete set of eigenfunctions with the equidistant spectrum:
\begin{align}
  &\mathcal{H}_{\ell}(g)\phi_{\ell,n}(x;g)
  =\mathcal{E}_{\ell,n}(g)\phi_{\ell,n}(x;g),\quad n=0,1,\ldots,\\[1pt]
  &\mathcal{E}_{\ell,n}(g)=\mathcal{E}_n(g+\ell)=4n,\\[-3pt]
  &\phi_{\ell,n}(x;g)=P_{\ell,n}(x^2;g)\psi_{\ell}(x),
  \,\psi_{\ell}(x)\eqdef\frac{e^{w_0(x;g+\ell)}}{\xi_{\ell}(x^2;g)},\\[-2pt]
  &P_{\ell,n}(x;g)\eqdef
  \xi_{\ell}(x;g+1)P_n(x;g+\ell)\n
  &\phantom{P_{\ell,n}(x;g)\eqdef}
  -\xi_{\ell-1}(x;g+2)P_{n-1}(x;g+\ell).
  \label{poleiglag}
\end{align}
Obviously we have $P_{\ell,0}(x;g)=\xi_{\ell}(x;g+1)$ and
$\phi_{\ell,0}(x;g)$ $=e^{w_{\ell}(x;g)}$.
The polynomial eigenfunction $P_{\ell,n}(x^2;g)$ is a degree $\ell+n$
polynomial in $x^2$ but it has only $n$ zeros in the domain $0<x<\infty$.
These polynomials are orthogonal with respect to the measure
$\psi_{\ell}(x;g)^2$:
\begin{align}
  &\int_0^{\infty}\!dx\,\psi_{\ell}(x;g)^2
  P_{\ell,n}\bigl(x^2;g\bigr)P_{\ell,m}\bigl(x^2;g\bigr)\n[-2pt]
  &\qquad
  =\frac{1}{2\,n!}\bigl(n+g+2\ell-\tfrac12\bigr)
  \Gamma\bigl(n+g+\ell-\tfrac12\bigr)\delta_{nm}.
\end{align}
They form a complete basis of the Hilbert space just like the
Laguerre polynomials in the $\ell=0$ case.
These new types of polynomials do not satisfy the three term recurrence
relation, a characteristic feature of all the ordinary orthogonal
polynomials.
It should be stressed that all four terms in \eqref{poleiglag} are
the Laguerre polynomials of the same index, $g+\ell-1/2$.
The action of the operators $\mathcal{A}_{\ell}(g)$ and
$\mathcal{A}_{\ell}(g)^{\dagger}$ on the eigenfunctions are:
\begin{align}
  &\mathcal{A}_{\ell}(g)\phi_{\ell,n}(x;g)=-2\phi_{\ell,n-1}(x;g+1),\n
  &\mathcal{A}_{\ell}(g)^{\dagger}\phi_{\ell,n-1}(x;g+1)
  =-2n\phi_{\ell,n}(x;g).
  \label{AAdonLag}
\end{align}

For $\ell=1$ the Hamiltonian reads
\begin{align}
  \mathcal{H}_1(g)&=p^2+
  x^2+\frac{g(g+1)}{x^2}-3-2 g\n[-2pt]
  &\quad +\frac{4}{x^2+g+\frac12}
  -\frac{4 (2 g+1)}{\left(x^2+g+\frac12\right)^2},
  \nonumber
\end{align}
which is equivalent to that of the shape invariant potential of
Quesne eq.(8) of \cite{quesne} with the replacement $\omega\to2$
and $l\to g$. The formula \eqref{poleiglag} expressing the polynomial
eigenfunctions in terms of the Laguerre polynomials is the
generalisation of G\'omez-Ullate et al.'s \cite{gomez} relation
eq.(80) between the $X_1$-Laguerre and the Laguerre polynomials.

\section{Darboux-P\"oschl-Teller potential}
\label{DPT}

Here we present another infinite number of shape invariant potentials
indexed by a non-negative integer $\ell=0,1,2,\ldots$.
For $\ell=0$, it is known as the P\"oschl-Teller potential \cite{PT},
with two positive parameters $\bm{\lambda}=(g,h)$:
\begin{align}
  &\mathcal{H}_0(\bm{\lambda})=p^2+\frac{g(g-1)}{\sin^2 x}
  +\frac{h(h-1)}{\cos^2 x}-(g+h)^2,\\
  &w_0(x;\bm{\lambda})=g\log\sin x+h\log\cos x,\quad 0<x<\frac{\pi}{2}.
\end{align}
We again follow the notation of our previous work \cite{os7} \S{III.A.2}.
The shape invariance, the Heisenberg operator solution and the
creation-annihilation operators of the above Hamiltonian are
discussed in some detail there.
Apparently, it was Darboux \cite{darboux} who first introduced this
potential, although the coupling constants were restricted to positive
integers only.
It is trivial to verify the shape invariance condition \eqref{shapeinv2}
with $\bm{\delta}=(1,1)$, $\mathcal{E}_1(\bm{\lambda})=4(1+g+h)$
and we obtain the quadratic energy spectrum and the corresponding
eigenfunctions $n=0,1,\ldots$,
\begin{align}
  &\mathcal{E}_n(\bm{\lambda})=4n(n+g+h),\\
  &\phi_n(x;\bm{\lambda})
  =P_n(\cos 2x;\bm{\lambda})\,e^{w_0(x;\bm{\lambda})},\n
  &\qquad
  P_n(x;\bm{\lambda})=P_n^{(g-\frac12,h-\frac12)}(x).
  \label{jacpoly}
\end{align}
The polynomial eigenfunctions are the Jacobi polynomials in $\cos2x$,
which are orthogonal with respect to the measure
$\phi_0(x;\bm{\lambda})^2=e^{2w_0(x;\bm{\lambda})}
=(\sin x)^{2g}(\cos x)^{2h}$.

For each positive integer $\ell\ge1$, let us introduce a prepotential
and a Hamiltonian \eqref{Aldef}--\eqref{Hldef} together with the
restriction on the parameters, $h>g>0$:
\begin{align}
  &\xi_{\ell}(x;\bm{\lambda})\eqdef
  P_{\ell}^{(-g-\ell-\frac12,h+\ell-\frac32)}(x),
  \label{xijac}\\
  &w_{\ell}(x;\bm{\lambda})\eqdef w_0(x;\bm{\lambda}+\ell\bm{\delta})
  +\log\frac{\xi_{\ell}(\cos 2x;\bm{\lambda}+\bm{\delta})}
  {\xi_{\ell}(\cos 2x;\bm{\lambda})}.
  \label{wljac}
\end{align}
Since the polynomial $\xi_{\ell}(\cos 2x;\bm{\lambda})$ has no zero in the
domain $0<x<\frac{\pi}{2}$, the prepotential and the potential are
smooth in the entire domain.
It is straightforward to verify \eqref{shapeinv2} with $\bm{\delta}=(1,1)$,
$\mathcal{E}_{\ell,1}(\bm{\lambda})=4(1+2\ell+g+h)$.
The eigenvalues and the eigenfunctions of
$\mathcal{H}_{\ell}(\bm{\lambda})$ have the following form:
\begin{align}
  &\mathcal{H}_{\ell}(\bm{\lambda})\phi_{\ell,n}(x;\bm{\lambda})
  =\mathcal{E}_{\ell,n}(\bm{\lambda})\phi_{\ell,n}(x;\bm{\lambda}),
  \  n=0,1,\ldots,
  \label{hptschr}\\
  &\mathcal{E}_{\ell,n}(\bm{\lambda})=
  \mathcal{E}_n(\bm{\lambda}+\ell\bm{\delta})=4n(n+2\ell+g+h),\\
  &\phi_{\ell,n}(x;\bm{\lambda})
  =P_{\ell,n}(\cos 2x;\bm{\lambda})\psi_{\ell}(x;\bm{\lambda}),
  \label{genformjac}\\
  &\psi_{\ell}(x;\bm{\lambda})\eqdef
  \frac{e^{w_0(x;\bm{\lambda}+\ell\bm{\delta})}}
  {\xi_{\ell}(\cos 2x;\bm{\lambda})},\\
  &P_{\ell,n}(x;\bm{\lambda})\eqdef
  a_{\ell,n}(x;\bm{\lambda})P_n(x;\bm{\lambda}+\ell\bm{\delta})\n
  &\phantom{P_{\ell,n}(x;\bm{\lambda})=}
  +b_{\ell,n}(x;\bm{\lambda})P_{n-1}(x;\bm{\lambda}+\ell\bm{\delta}),
  \label{Plnjac}\\
  &a_{\ell,n}(x;\bm{\lambda})\eqdef\xi_{\ell}(x;g+1,h+1)\n[-2pt]
  &\quad
  +\frac{2n(-g+h+\ell-1)\,\xi_{\ell-1}(x;g,h+2)}
  {(-g+h+2\ell-2)(g+h+2n+2\ell-1)}\n[-1pt]
  &\quad-\frac{n(2h+4\ell-3)\,\xi_{\ell-2}(x;g+1,h+3)}
  {(2g+2n+1)(-g+h+2\ell-2)},\\
  &b_{\ell,n}(x;\bm{\lambda})\eqdef
  \frac{(-g+h+\ell-1)(2g+2n+2\ell-1)}{(2g+2n+1)(g+h+2n+2\ell-1)}\n
  &\qquad\qquad\qquad\times
  \xi_{\ell-1}(x;g,h+2).
\end{align}
The polynomial eigenfunction $P_{\ell,n}(x;\bm{\lambda})$ is a degree
$\ell+n$ polynomial in $x$ and we have $P_{\ell,0}(x;\bm{\lambda})
=\xi_{\ell}(x;\bm{\lambda}+\bm{\delta})$ and
$\phi_{\ell,0}(x;\bm{\lambda})=e^{w_{\ell}(x;\bm{\lambda})}$.
Again $P_{\ell,n}(\cos 2x;\bm{\lambda})$ has only $n$ zeros
in the domain $0<x<\frac{\pi}{2}$.
It should be stressed that $P_{\ell,n}(x;\bm{\lambda})$ are
polynomials in the coupling constants $g,h$.
They are orthogonal with respect to the measure
$\psi_{\ell}(x;\bm{\lambda})^2$,
\begin{align}
  &\int_0^{\frac{\pi}{2}}\!dx\,\psi_{\ell}(x;\bm{\lambda})^2
  P_{\ell,n}\bigl(\cos 2x;\bm{\lambda}\bigr)
  P_{\ell,m}\bigl(\cos 2x;\bm{\lambda}\bigr)\n[-2pt]
  &=\frac{\Gamma(n+g+\ell+\frac12)\Gamma(n+h+\ell+\frac12)}
  {2\,n!\,(2n+g+h+2\ell)\Gamma(n+g+h+2\ell)}\n
  &\quad\times
  \frac{(n+g+\ell+\frac12)(n+h+2\ell-\frac12)}
  {(n+g+\frac12)(n+h+\ell-\frac12)}\,\delta_{nm},
\end{align}
and they form a complete basis of the Hilbert space just like the
Jacobi polynomials in the $\ell=0$ case.
The action of the operators $\mathcal{A}_{\ell}(\bm{\lambda})$ and
$\mathcal{A}_{\ell}(\bm{\lambda})^{\dagger}$ on the eigenfunctions are:
\begin{align}
  &\mathcal{A}_{\ell}(\bm{\lambda})\phi_{\ell,n}(x;\bm{\lambda})
  =-2(n+2\ell+g+h)\phi_{\ell,n-1}(x;\bm{\lambda}+\bm{\delta}),\n
  &\mathcal{A}_{\ell}(\bm{\lambda})^{\dagger}
  \phi_{\ell,n-1}(x;\bm{\lambda}+\bm{\delta})
  =-2n\phi_{\ell,n}(x;\bm{\lambda}).
  \label{AAdonJac}
\end{align}

For $\ell=1$ the Hamiltonian reads explicitly as
\begin{align}
  &\mathcal{H}_1(\bm{\lambda})=p^2+\frac{g(g+1)}{\sin^2 x}
  +\frac{h(h+1)}{\cos^2 x}-(2+g+h)^2\n
  &+\frac{8(g+h+1)}{1\!+\!g\!+\!h\!+\!(g-\!h)\cos2x}
  -\frac{8(2g+1)(2h+1)}{(1\!+\!g\!+\!h\!+\!(g-\!h)\cos2x)^2},
  \nonumber
\end{align}
which is equivalent to that of the shape invariant potential of
Quesne eq.(11) of \cite{quesne} with the replacement $A\to\frac12(g+h)+1$,
$B\to\frac12(h-g)$ and $x\to 2(\frac{\pi}{4}-x)$.
The formula \eqref{Plnjac} expressing the polynomial eigenfunctions
in terms of the Jacobi polynomials is the generalisation of
G\'omez-Ullate et al.'s \cite{gomez} relation eq.(56) between
the $X_1$-Jacobi and the Jacobi polynomials.

\section{Hyperbolic -P\"oschl-Teller potential}
\label{HPT}

The next example provides only a finite number of shape invariant
potentials, as many as the existing bound states of the starting
Hamiltonian with the hyperbolic P\"oschl-Teller potential with
$\bm{\lambda}=(g,h)$, $h>g>0$:
\begin{align}
  &\mathcal{H}_0(\bm{\lambda})=p^2+\frac{g(g-1)}{\sinh^2 x}
  -\frac{h(h+1)}{\cosh^2 x}+(h-g)^2,\\
  &w_0(x;\bm{\lambda})=g\log\sinh x-h\log\cosh x, \ 0<x<\infty.
\end{align}
As the name suggests, it is the hyperbolic function version of the
Darboux-P\"oschl-Teller model discussed in the preceding section.
It is trivial to verify 
\eqref{shapeinv2} with  $\bm{\delta}=(1,-1)$,
$\mathcal{E}_1(\bm{\lambda})=4(h-g-1)$.
We obtain the quadratic energy spectrum and the corresponding
eigenfunctions $n=0,1,\ldots, n_{\!B}\eqdef[(h-g)/2]'$, expressed in terms
of the Jacobi polynomials:
\begin{align}
  &\mathcal{E}_n(\bm{\lambda})=4n(h-g-n), \quad n=0,1,2,\ldots,n_{\!B},\\
  &\phi_n(x;\bm{\lambda})=P_n(\cosh 2x;\bm{\lambda})
  \,e^{w_0(x;\bm{\lambda})},\n
  &\qquad
  P_n(x;\bm{\lambda})=P_n^{(g-\frac12,-h-\frac12)}(x).
\end{align}
Here $[x]'$ denotes the greatest integer not equal or exceeding $x$.
These finite number of polynomials in $\cosh2x$ are square integrable
and are orthogonal with respect to the measure
$\phi_0(x;\bm{\lambda})^2=e^{2w_0(x;\bm{\lambda})}=(\sinh x)^{2g}
(\cosh x)^{-2h}$.

For each positive integer $1\le\ell< n_{\!B}$, let us introduce a
prepotential and a Hamiltonian \eqref{Aldef}--\eqref{Hldef}:
\begin{align}
  &\xi_{\ell}(x;\bm{\lambda})\eqdef
  P_{\ell}^{(-g-\ell-\frac12,-h+\ell-\frac32)}(x),\\
  &w_{\ell}(x;\bm{\lambda})\eqdef w_0(x;\bm{\lambda}+\ell\bm{\delta})
  +\log\frac{\xi_{\ell}(\cosh 2x;\bm{\lambda}+\bm{\delta})}
  {\xi_{\ell}(\cosh 2x;\bm{\lambda})}.\!\!
\end{align}
Since the polynomial $\xi_{\ell}(\cosh 2x;\bm{\lambda})$,
$1\le \ell< n_{\!B}$, has no zero in the domain $0<x<\infty$,
the prepotential and the potential are smooth in the entire domain.
It is straightforward to verify \eqref{shapeinv2} with
$\bm{\delta}=(1,-1)$,
$\mathcal{E}_{\ell,1}(\bm{\lambda})=4(h-g-2\ell-1)$.
The repetition of the shape invariant transformation
$w_{\ell}(x;\bm{\lambda})\to w_{\ell}(x;\bm{\lambda}+\bm{\delta})$ is
unlimited for the previous two examples.
For the hyperbolic P\"oschl-Teller Hamiltonian
$\mathcal{H}_{\ell}(\bm{\lambda})$, the maximal repetition is $n_{\!B}-\ell$.
Beyond that point, the transformed Hamiltonian no longer possesses
a bound state. The maximal $\ell$ case, $\mathcal{H}_{n_{\!B}}$ has only
one bound state, which is exactly calculable, and the transformed
one has none.
The energy spectrum is
\begin{align}
  &\mathcal{E}_{\ell,n}(\bm{\lambda})
  =\mathcal{E}_n(\bm{\lambda}+\ell\bm{\delta})=4n(h-g-2\ell-n),\n
  &\qquad\qquad\qquad
  n=0,\ldots, n_{\!B}-\ell.
\end{align}
Using the same notation as \eqref{hptschr} and
\eqref{genformjac}--\eqref{Plnjac} with the replacement
$\cos 2x\to\cosh 2x$, the eigenfunctions are
\begin{align}
  &a_{\ell,n}(x;\bm{\lambda})\eqdef\xi_{\ell}(x;g+1,h-1)\n[-2pt]
  &\quad
  +\frac{2n(-g-h+\ell-1)\,\xi_{\ell-1}(x;g,h-2)}
  {(-g-h+2\ell-2)(g-h+2n+2\ell-1)}\n[-1pt]
  &\quad-\frac{n(-2h+4\ell-3)\,\xi_{\ell-2}(x;g+1,h-3)}
  {(2g+2n+1)(-g-h+2\ell-2)},\\
  &b_{\ell,n}(x;\bm{\lambda})\eqdef
  \frac{(-g-h+\ell-1)(2g+2n+2\ell-1)}{(2g+2n+1)(g-h+2n+2\ell-1)}\n
  &\qquad\qquad\qquad\times
  \xi_{\ell-1}(x;g,h-2).
\end{align}
The polynomials $\{P_{\ell,n}(x;\bm{\lambda})\}$ are orthogonal
with respect to the measure $\psi_{\ell}(x;\bm{\lambda})^2$.
The action of the operators $\mathcal{A}_{\ell}(\bm{\lambda})$ and
$\mathcal{A}_{\ell}(\bm{\lambda})^{\dagger}$ on the eigenfunctions are:
\begin{align}
  &\mathcal{A}_{\ell}(\bm{\lambda})\phi_{\ell,n}(x;\bm{\lambda})
  =-2(h-g-2l-n)\phi_{\ell,n-1}(x;\bm{\lambda}+\bm{\delta}),\n
  &\mathcal{A}_{\ell}(\bm{\lambda})^{\dagger}
  \phi_{\ell,n-1}(x;\bm{\lambda}+\bm{\delta})
  =-2n\phi_{\ell,n}(x;\bm{\lambda}).
  \label{AAdonhJac}
\end{align}
The $\ell=1$ Hamiltonian reads
\begin{align}
  &\mathcal{H}_1(\bm{\lambda})=p^2+\frac{g(g+1)}{\sinh^2 x}
  -\frac{h(h-1)}{\cosh^2 x}+(h-g-2)^2
  \label{bcr}\\
  &\!+\!\frac{8(h-g-1)}{1\!+\!g\!-\!h\!+\!(g+\!h)\!\cosh2x}
  \!-\!\frac{8(2g+1)(2h-1)}{(1\!+\!g\!-\!h\!+\!(g+\!h)\!\cosh2x)^2}.
  \nonumber
\end{align}

\section{Summary and Comments}
\label{comments}

By deforming two well-known shape invariant potentials in terms of
their eigenpolynomials, two infinite sets of shape invariant
potentials are obtained. Their eigenpolynomials form new types of
orthogonal polynomials starting with degree $\ell$, which is the
degree of the polynomial used for deformation.
It would be interesting to try to deform other shape invariant
potentials in a similar way.
It is a good challenge to clarify various properties of these new
polynomials, {\em e.g.\/} 
generating functions, the Gram-Schmidt construction, substitutes of
the three term recurrence relations, etc., and to pursue possible
physical applications.

The forward shift operator $\mathcal{F}_{\ell}$ and the backward shift
operator $\mathcal{B}_{\ell}$ are defined by
\begin{align}
   \mathcal{F}_{\ell}(\bm{\lambda})&\eqdef
   \psi_{\ell}(x\,;\bm{\lambda}+\bm{\delta})^{-1}\circ
   \mathcal{A}_{\ell}(\bm{\lambda})\circ\psi_{\ell}(x\,;\bm{\lambda}),\n
   \mathcal{B}_{\ell}(\bm{\lambda})&\eqdef
   \psi_{\ell}(x\,;\bm{\lambda})^{-1}\circ
   \mathcal{A}_{\ell}(\bm{\lambda})^{\dagger}
   \circ\psi_{\ell}(x\,;\bm{\lambda}+\bm{\delta}),
\end{align}
and their action on the eigenpolynomials $P_{\ell,n}$
can be read from \eqref{AAdonLag}, \eqref{AAdonJac} and \eqref{AAdonhJac}.

\medskip

The $\ell=1$ Hamiltonian $\mathcal{H}_{1}(\bm{\lambda})$ \eqref{bcr}
for the deformed hyperbolic P\"oschl-Teller potential is equivalent to
the `new extended potential' (9) of Bagchi et al.'s paper \cite{BQR}
with the replacement $A\to\frac12(h-g)-1$, $B\to\frac12(h+g)$ and $x\to 2x$.
We thank C. Quesne for pointing this out.
After submitting the present Letter for publication, a new paper appeared
\cite{quesne2}, which discussed the $\ell=2$ deformed Hamiltonian
$\mathcal{H}_{2}(\bm{\lambda})$ \eqref{Aldef}--\eqref{Hldef} with
\eqref{xilag}--\eqref{wllag} or \eqref{xijac}--\eqref{wljac}, and other
potentials related to the $X_2$-Laguerre or $X_2$-Jacobi polynomials.

\bigskip
This work is supported in part by Grants-in-Aid for Scientific Research
from the Ministry of Education, Culture, Sports, Science and Technology,
No.19540179.


\end{document}